\documentclass[useAMS,usenatbib]{mn2e}

\usepackage{graphicx}
\usepackage{natbib}

\title[The Mass Function of $\omega$ Cen] {The Mass Function of $\omega$ Centauri down to 0.15
$M_{\odot}$\thanks{Based on 
observations collected at the European Southern Observatory
within the observing program 74.D-0369(B).Also based on observations with the NASA/ESA Hubble 
Space Telescope, obtained at the Space 
Telescope Science Institute, which is operated by the Association of Universities for Research 
in Astronomy, Inc., under NASA contract NAS5-26555.}}
\author[Sollima et al.]{A. Sollima$^{1}$\thanks{E-mail:
antonio.sollima@bo.astro.it (AS)}, F.R. Ferraro$^{1}$ and M. Bellazzini$^{2}$\\
$^{1}$Dipartimento di Astronomia, Universit\`a di Bologna, via Ranzani 1,
Bologna, 40127-I, Italy\\
$^{2}$INAF Osservatorio Astronomico di Bologna, via Ranzani 1,
Bologna, 40127-I, Italy}
\begin{document}

\date{Accepted 2006, ???; Received 2006, ???; in original form
2006, ???}

\pagerange{\pageref{firstpage}--\pageref{lastpage}} \pubyear{2006}

\maketitle

\label{firstpage}

\begin{abstract}
By means of deep FORS1@VLT and ACS@HST observations of a wide area in 
the stellar system $\omega$ Cen we measured the luminosity function of main
sequence stars down to R=22.6 and $I_{F814W}$=24.5 . 
The luminosity functions obtained have been converted into mass functions and 
compared with analytical Initial Mass Functions (IMFs) available in the literature.
The mass function obtained, reaching $M\sim0.15~M_{\odot}$, can be well reproduced by a broken power-law 
with indices $\alpha=-2.3$ for $M>0.5M_{\odot}$ and $\alpha=-0.8$ for $M<0.5M_{\odot}$.
Since the stellar populations of $\omega$ Cen have been proved to be actually unaffected by dynamical evolution 
processes, the mass function measured in this stellar system should represent the best approximation of the IMF 
of a star cluster. The comparison with the MF measured in other Galactic 
globular clusters suggests that possible primordial differences in the slope of the low-mass end of their MF could 
exist. 
  
\end{abstract}


\section{Introduction}

The determination of the stellar Initial Mass Function (IMF) still represents
one of the most crucial questions in astrophysics. In fact, it is a critical 
ingredient in the understanding of a large number of basic astronomical 
phenomena such as the formation of the first stars, galaxy formation and 
evolution, and the determination of the possible dark matter content of galaxy 
halos.
After the pioneering study by
Salpeter (1955), a number of works have been carried out with the aim of
studying the shape of the IMF in the solar neighborhood (Miller \& Scalo 1979; 
Larson 1998; Chabrier 2001; Kroupa 2002). 

Star clusters provides a useful tool to
investigate the low-mass end of the IMF. They offer the possibility of observing
large samples of unevolved low-mass stars that are coeval and at the same distance with 
the same chemical composition. 
However, the derivation of the IMF in star clusters is complicated by the
tidal interaction with the Galaxy that drives star clusters preferentially to lose 
low-mass stars across the tidal boundary as a result of the ever-continuing 
redistribution of energy on the two-body relaxation time-scale.
De Marchi et al. (2005) found that the
luminosity function (LF) of Galactic globular clusters (GC) is well 
reproduced by adopting a Mass Function (MF) in which the number of stars per unit mass 
decreases below a critical mass.
Table 1 lists the most used analytical descriptions of the IMF available in
literature. A number of other power-law indices have been measured in stellar associations and young
star clusters in the local Universe (see Kroupa 2002 for a recent review).

The stellar system $\omega$ Centauri (NGC 5139) is the most massive and luminous
GC of the Milky Way (M$\sim2.5~10^{6}~M_{\odot}$, Van de Ven et
al. 2006) and plays a key role in the understanding of the properties of the 
mass spectrum for low-mass stars.
Ferraro et al. (2006) showed that the most massive objects in the cluster (blue 
straggler stars and interacting binaries) are not centrally segregated, at odds with any other 
GC. This evidence suggests that $\omega$ Cen is still
far from being completely relaxed even in the core region.
For this reason its present-day MF should reflect the IMF much more 
closely than in any other Galactic GC.

In this paper we measured the LF of $\omega$ Cen by means of
wide-field ground-based photometry using FORS1 and ACS@HST observations of a
peripheral region of the cluster. The obtained LFs have been compared with the LFs of other
Galactic GCs and converted into mass function (MF). 
In \S 2 we describe the observational material. \S 3 is devoted to the description of 
the procedure applied to derive the LFs. In \S 4 the derived LF is compared with those of a sample of Galactic GCs. 
In \S 5 the derived MF of $\omega$ Cen is presented and
compared with the analytical IMFs available in the literature. We discuss our results in \S 6.

\begin{table*}
\label{imf}
\caption{Summary of the most common analytical IMF in the literature}
\begin{tabular}{@{}lll@{}}
\hline
Salpeter (1955)        & $A~m^{\alpha}$ 				 & $\alpha=-2.35$ \\
Miller \& Scalo (1979) & $\frac{A}{m}~e^{-\frac{(log~m- log~m_{0})^{2}}{2~\sigma_{log~m}^{2}}}$ &
$log~m_{0}=-1.02$ \\
                       &                                                   & $\sigma_{log~m}=0.68$\\
Larson (1998)(a)       & $A~m^{\alpha}~e^{-\frac{m_{0}}{m}}$		   & $\alpha=-2.35$\\
                       &  						   & $m_{0}=0.3 M_{\odot}$\\
Larson (1998)(b)       & $A~m^{\alpha}~(1-e^{-\frac{m}{m_{0}}})$           & $\alpha=-2.35$\\
                       &                                                   & $m_{0}=0.35 M_{\odot}$\\						  
Chabrier (2001)        & $A~m^{-\delta}e^{-\frac{m_{0}}{m}^{\beta}}$	   & $\delta=3.3$ \\
                       &                                                   & $m_{0}=716.4 M_{\odot}$\\
Kroupa (2002)          & $A~(\frac{m}{m_{0}})^{\alpha}$                    & $m_{0}=0.08 M_{\odot}~;\alpha=-0.3~~for~0.01M_{\odot}<M<0.08~M_{\odot}$\\
                       &                                                   & $m_{0}=0.08 M_{\odot}~;\alpha=-1.3~~for~0.08M_{\odot}<M<0.5~M_{\odot}$\\
                       &                                                   & $m_{0}=0.5 M_{\odot}~;\alpha=-2.3~~for~0.5M_{\odot}<M<1~M_{\odot}$\\
\hline
\end{tabular}
\end{table*}

\section{Observations}
\label{redums}

The analysis presented here is based on two photometric datasets: {\it (i)} A mosaic of
eight deep images obtained with FORS1@VLT in the B and R passbands and {\it (ii)} a set 
of high-resolution images obtained with ACS@HST through the F606W and F814W filters
in an external region of the cluster.

\begin{figure}
 \includegraphics[width=8.7cm]{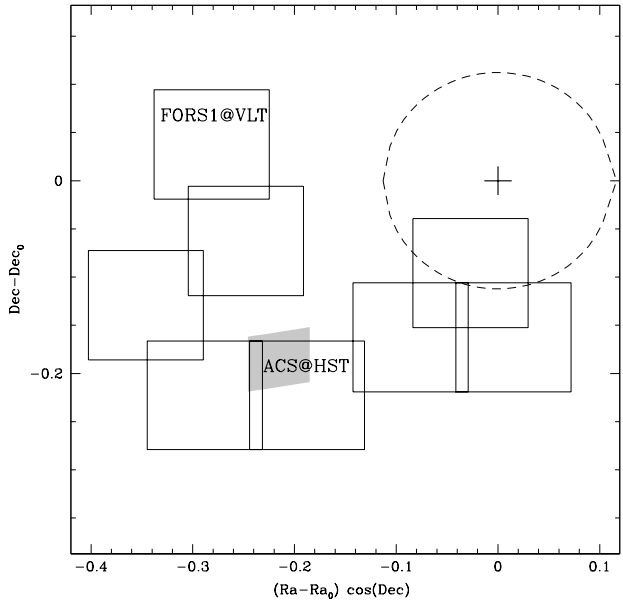}
\caption{Map of the region sampled by the FORS1 observations. North is up, East on
the right. The eight fields observed with FORS1 are shown. The grey box indicates the position of
the ACS field. The cluster center and half-mass radius are indicated by the black cross and the 
dashed line, respectively.}
\label{map}
\end{figure}

FORS1 observations sample the Main Sequence (MS) population of $\omega$ Cen down 
to $R\sim24$ containing more than 70,000 stars between 6' and 27' ($\sim 10 r_{c}$) from the cluster
center. The photometry has been performed with the PSF-fitting code 
DoPhot (Schechter et al. 1993). A detailed description of the data reduction and calibration procedure
can be found in Sollima et al. (2007).
The outermost field of the FORS1 observations partially overlaps the deep ACS 
photometry allowing one to link these two datasets (see Fig. \ref{map}).

ACS observations cover a small region of $3.4'\times~3.4'$ in an external field
at $\sim 17'$ from the cluster center. They consists in a set of four 1300 s
and 1340 s long exposures through the F606W and F814W filters, respectively. 
The photometric analysis have been performed using the
SExtractor photometric package (Bertin \& Arnouts 1996). Given the small stellar 
density in this field ($\sim 0.1~stars/arcsec^2$), crowding does not affect the aperture photometry, 
allowing to properly estimate the magnitude of stars. For each star we measured 
the flux contained within a radius of 0.125" ($\sim$ FWHM). 
The source detection and photometric analysis have been performed 
independently on each image. Only stars detected in three out of four frames have been 
included in the final catalog. The most isolated and
brightest stars in the field have been used to link the
0.125"- to  0.5"-aperture photometry, after normalizing for exposure time. 
Instrumental magnitudes have been transformed into
the VEGA-MAG system by using the photometric zero-points by Sirianni et al.
(2005). The obtained catalog contains 5,440 stars reaching the limiting magnitudes
$V_{F606W} \sim$ 27 and $I_{F814W} \sim$ 25.

\begin{figure}
 \includegraphics[width=8.7cm]{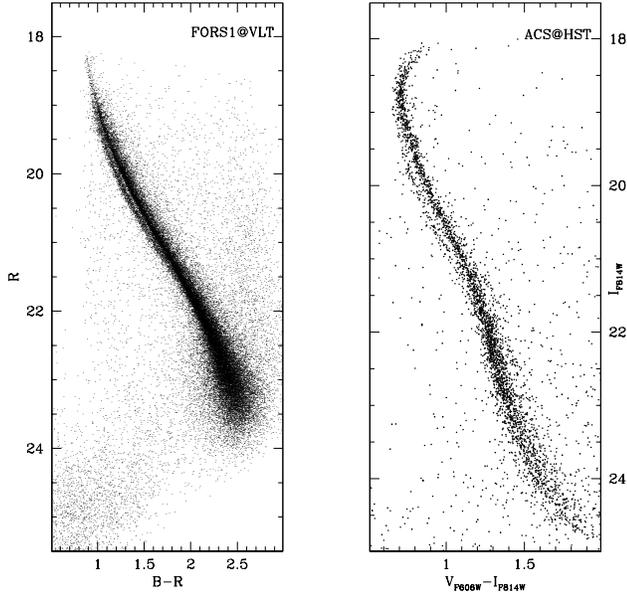}
\caption{FORS1 (R, B-R; left $panel$) and ACS ($I_{F814W}, V_{F606W}-I_{F814W}$; right $panel$) CMDs of $\omega$ Cen.}
\label{cmd}
\end{figure}

Fig. \ref{cmd} shows the color-magnitude diagrams (CMDs) obtained for the two 
samples. The main features of the presented CMDs are
schematically listed below:
\begin{itemize}
\item{The CMDs sample the unevolved population of $\omega$ Cen from the turn-off
down to the lower MS;}
\item{A narrow blue MS (bMS), running parallel to the dominant MS population, can be
distinguished at $19.5<R<21$ and $19<I_{F814W}<21$ . This feature has been already 
described and discussed in Bedin et al. (2004) and Sollima et al. (2007);}
\item{ACS observations reach a fainter limiting magnitude, sampling also the 
region close to the hydrogen burning limit.}
\end{itemize}

In the following section we describe the adopted procedure to derive the
LFs and the global MF of $\omega$ Cen from these 
data-sets.
  
\section{Luminosity Function}
\label{method}

To compute the LF of $\omega$ Cen we followed the procedure described below.
As a first step, we computed the MS ridge line by averaging the colors of stars in the 
CMD over 0.2 mag boxes and applying a 2$\sigma$ clipping algorithm.
Then, the LF has been computed by counting the number of 
objects in 0.5 mag wide bins separated by 0.1 mag along the R and $I_{F814W}$ axes. 
Only stars within $\pm 2.5$ times the color 
standard deviation around the MS ridge line have been considered (see De Marchi \& Paresce 1995).

\begin{figure}
 \includegraphics[width=8.7cm]{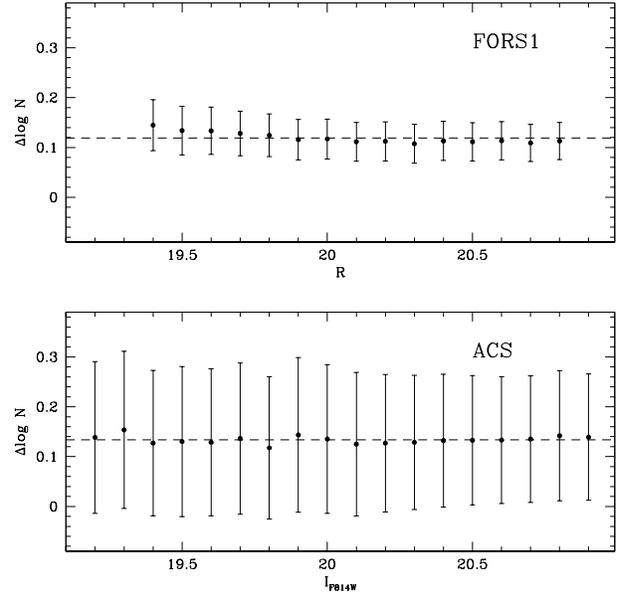}
\caption{Difference between the LFs calculated with and without bMS stars for the FORS1 
(upper $panel$) and the ACS (lower $panel$) sample. The average constant trend is indicated 
as a dashed line in both $panels$.}
\label{he}
\end{figure}

$\omega$ Cen is well known to harbour stellar populations with different metallicity 
(Norris, Freeman \& Mighell 1996 and references therein) and, possibly, helium content (Norris 2004).
The CMDs shown in Fig. \ref{cmd} do 
not allow one to distinguish the different MS components of the cluster (except for
the bMS population), making impossible to disentangle the contribution of each
population to the LF. However, the metal-poor population of $\omega$ Cen comprises
more than 70\% of the entire cluster content (Pancino et al. 2000), thus
dominating the shape of the LF.
To check the validity of this assumption, we measured the LF of MS stars by excluding 
bMS stars in the magnitude range where this sequence is clearly distinguishable from the 
dominant cluster population.
For this purpose, 
we considered bMS stars all the objects lying at a distance 
between 0.5 and 2.5 times  the color standard deviation around the 
cluster ridge line on the blue side of the dominant cluster MS in the magnitude range 
$19.2<R<20.7$ and $19.2<I_{F814W}<20.9$ for the FORS1 and ACS sample, respectively.
We found a constant difference between the LF measured with and without 
bMS stars over the entire magnitude range under investigation in both photometric samples.
In particular, in both samples the deviations from the constant difference lie within $\Delta log~N<0.03$
over the entire magnitude range considered here (see Fig. \ref{he}).
This indicates that although bMS stars constitute a significant fraction of the cluster population 
($\sim$ 24\%, Sollima et al. 2007), their presence does not significantly distort the shape of the 
measured LF.
For this reason, in the following we will measure the LF of the global MS population assuming 
it as representative of the dominant cluster population.

The LF has been calculated by
taking into account two important effects: {\it (i)} the photometric incompleteness and
{\it (ii)} the contamination from field stars. In the following sections the adopted techniques
to correct for these effects are described.

\subsection{Photometry Incompleteness}

A reliable determination 
of the LF from the CMDs of Fig. \ref{cmd} requires the assessment of the degree of 
photometric incompleteness as a function of magnitude and color.
For each individual field, the adopted procedure for artificial star experiments has been
performed as follows (see Bellazzini et al. 2002):
\begin{itemize}
\item{The magnitude of artificial stars was randomly extracted from a LF modeled to 
reproduce the observed LF for bright stars ($R < 22$; $I_{F814W} < 22$) and to provide large 
numbers of faint 
stars down to below the detection limits of our observations ($R \sim 24$; $I_{F814W} \sim 25$)\footnote{Note 
that the assumption for the fainter stars is only for statistical purposes, i.e., to 
simulate a large number of stars in the range of magnitude where significant losses, due to 
incompleteness, are expected.}. The color of each star was obtained by deriving, for each 
extracted R and $I_{F814W}$ magnitude, the corresponding B and $V_{F606W}$ magnitude, for the two datasets
respectively, by interpolating on the 
cluster ridge line. Thus, all the artificial stars lie on the cluster ridge line in the CMD;}
\item{We divided the frames into grids of cells of known width (30 pixels) and randomly positioned 
only one artificial star per cell for each run\footnote{We constrain each artificial star to 
have a minimum distance (5 pixels) from the edges of the cell. In this way we can control the 
minimum distance between adjacent artificial stars. At each run the absolute position of the 
grid is randomly changed in a way that, after a large number of experiments, the stars are 
uniformly distributed in coordinates.};}
\item{For the FORS1 sample, artificial stars were simulated with the DoPhot (Schechter et al. 1993)
model for the fit, including any spatial 
variation of the shape of the PSF. For the ACS sample, artificial stars were
simulated as gaussians with a $FWHM=0.1"$. Artificial stars were added on the original frames 
including Poisson photon noise. Each star was added to both B and R (F606W and F814W for the ACS
sample) frames. The measurement process was 
repeated adopting the same procedure of the original measures and applying the same 
selection criteria described in Sollima et al. (2007) and in Sect. \ref{redums};}
\item{The results of each single set of simulations were appended to a file until the 
desired total number of artificial stars was reached. The final result for each sub-field is a 
list containing the input and output values of positions and magnitudes.}
\end{itemize}

\begin{figure}
 \includegraphics[width=8.7cm]{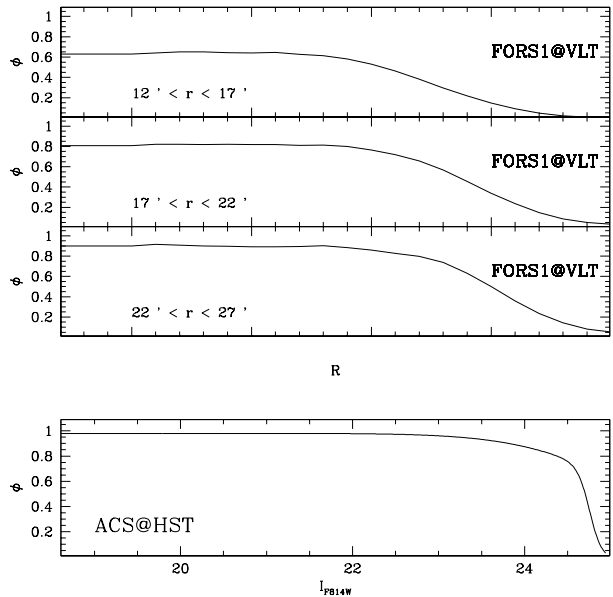}
\caption{Completeness factor $\phi$ as a function of R magnitude at
three different distances from the cluster center for the FORS1 sample (top three $panels$) and as
a function of $I_{F814W}$ magnitude for the ACS sample (bottom $panel$).}
\label{compl}
\end{figure}

More than 100,000 artificial stars have been produced providing a robust estimate of the
photometric completeness over the entire magnitude extension of the MS. 
Fig. \ref{compl} shows the completeness factor ($\phi$) as a function of the R magnitude at
three different distances from the cluster center for the FORS1 sample and as a function of the $I_F814W$ 
magnitude for the ACS sample. 
Only stars lying in the magnitude ranges where $\phi > 0.5$ were used to construct the LF.
  
\subsection{Field Contamination}
\label{field}

The contamination due to field stars was 
taken into account by using the Galaxy model of Robin et al. (2003). A synthetic
catalog covering an area of 0.5 square degrees in the cluster direction has been retrieved.
A sub-sample of stars has been randomly extracted from the entire catalog scaled to the
ACS and FORS1 field of view.      
For the ACS sample, the V and I Johnson-Cousin magnitudes were converted into the ACS photometric system with the
transformations of Sirianni et al. (2005).
The number of stars contained in each magnitude bin within the color window used to measure the LF (see above) 
has been subtracted from the completeness-corrected MS star counts.
The density of field objects in each magnitude bin is rather low (of the order of $<$2\%), in 
agreement with theoretical expectations (M\'endez \& Van Altena 1996).

\subsection{Results}
\label{resforsms}

\begin{figure}
 \includegraphics[width=8.7cm]{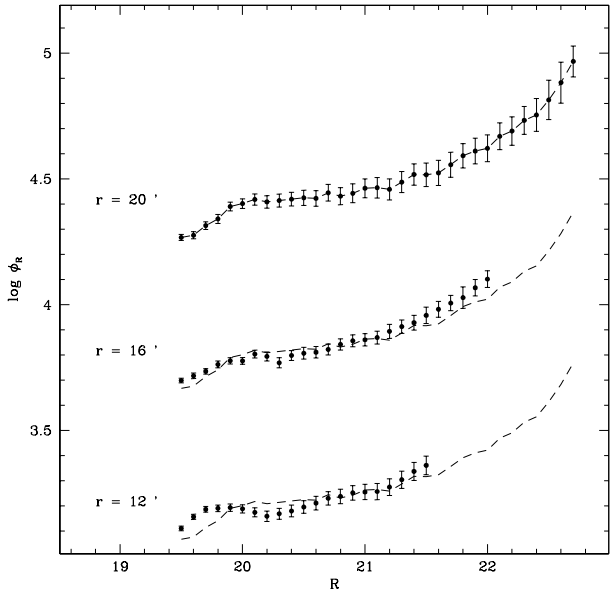}
\caption{R LF of $\omega$ Cen calculated between 12' and 20' from the cluster 
center. Each LF has been arbitrarily shifted of 0.15 in the y direction for clearity. 
The LF measured at 20' is indicated for comparison with dashed lines.}
\label{lfors}
\end{figure}

Fig. \ref{lfors} shows the R band LF of $\omega$ Cen calculated using stars lying in annuli of
4' width at different distances from the cluster center. 
Note that the $\phi=0.5$ limit is reached at brighter magnitudes as inner (more crowded) regions 
are considered.
As can be seen, in the magnitude
range covered in this analysis, no significant variations are visible in the shape of the LFs calculated between 12'
and 20' . This evidence provides further support to the hypothesis that this cluster is not affected by 
mass segregation.
This result is in agreement with the findings of Anderson (1997) who compared
the observed LF of $\omega$ Cen with the LF predicted by theoretical models 
with and without equipartition.
Strong evidence of the lack of equipartition in $\omega$ Cen has been provided by Ferraro et al.
(2006) on the basis of the comparison between the radial distribution of the blue 
straggler stars with that of the normal less massive cluster stars.
As a consequence, the LF measured here can be used to derive a MF which can be considered a good 
approximation of the cluster IMF. 

\begin{figure}
 \includegraphics[width=8.7cm]{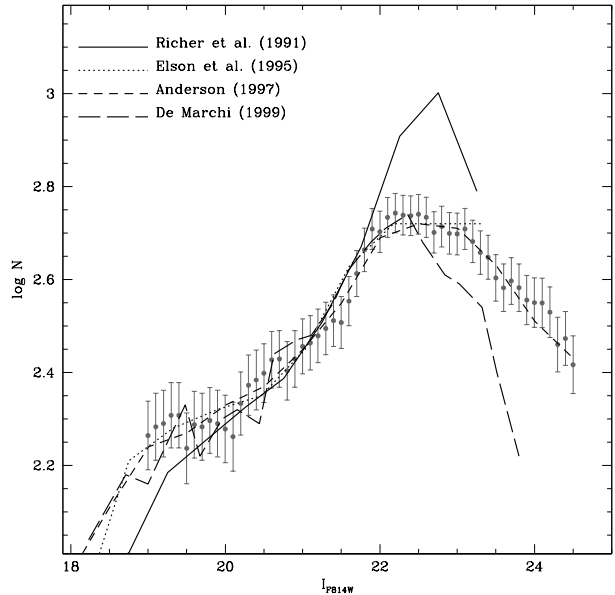}
\caption{F814W LF of $\omega$ Cen calculated from the ACS external field (grey points).
LFs by Richer et al. (1991, solid line), Elson et al. (1995, dotted line), Anderson (1997, short
dashed line) and De Marchi (1999, long dashed line) are overplotted.}
\label{lfacs}
\end{figure}

The F814W LF calculated for the external ACS field is shown in Fig. \ref{lfacs} and listed in
Table 2. The LF reaches a peak at $I_{F814W} \sim$22.2 and clearly drops at fainter magnitudes, well before
photometric incompleteness becomes significant. The LF of $\omega$ Cen measured by Richer et al. (1991), Elson et
al. (1995), Anderson (1997) and De Marchi (1999) are overplotted to our data in Fig. \ref{lfacs}. The original
magnitudes provided by these authors have been converted into the ACS F814W magnitude with the
transformations by Sirianni et al. (2005). All LFs showed in Fig. \ref{lfacs}
have been normalized to have the same number of stars in the magnitude range $20<I_{F814W}<22$ .
We note that the LF obtained in the present analysis
extends to fainter magnitudes than those by Richer et al. (1991) and Elson et al. (1995), having
a deepness comparable with those of De Marchi (1999) and Anderson (1997). Our ACS LF is
in good agreement with that of Elson et al. (1995) and Anderson (1997). The LF by Richer et al.
(1991) shows a steeper slope and continues to rise below $I_{F814W}>22.5$, at odds with the
other LFs. This discrepancy is likely to be due to uncertainties in the number counts and 
completeness corrections at the faint end of the LF by Richer et al. (1991) whose observations were obtained from 
ground-based observations where crowding effects can produce severe incompleteness. The LF by De Marchi
(1999) shows a drop for $I_{F814W}>22.5$ that is much more sudden than in our LF. However, the photometric 
dataset used by De Marchi (1999) has a limiting magnitude $\sim 1~mag$ brighter than that
reached by our observations and is four times less populous than the ACS sample. For these reasons,
we consider our LF more reliable at least at its faint end.
 
\begin{table}
\label{lf}
\caption{LF of $\omega$ Cen measured in the ACS field}
\begin{tabular}{@{}lllll@{}}
\hline
$I_{F814W}$ & $N_{obs}$ & $N_{field}$ & $\phi$ & N\\ 
\hline
19.0 & 183 & 4 & 0.980 & 183.7\\
19.1 & 190 & 6 & 0.980 & 190.9\\
19.2 & 194 & 7  & 0.980 & 195.0\\
19.3 & 202 & 7  & 0.980 & 202.1\\
19.4 & 202 & 5  & 0.980 & 202.1\\
19.5 & 170 & 3  & 0.980 & 172.5\\
19.6 & 191 & 3  & 0.980 & 191.9\\
19.7 & 189 & 3  & 0.980 & 189.9\\
19.8 & 194 & 2  & 0.980 & 194.0\\
19.9 & 192 & 2  & 0.980 & 193.9\\
20.0 & 187 & 3  & 0.980 & 186.8\\   
20.1 & 180 & 1  & 0.980 & 180.7\\
20.2 & 212 & 5  & 0.980 & 212.3\\
20.3 & 234 & 5 & 0.980 & 234.8\\
20.4 & 242 & 5 & 0.980 & 241.9\\
20.5 & 251 & 5 & 0.980 & 252.1\\
20.6 & 268 & 4 & 0.980 & 264.5\\
20.7 & 269 & 3 & 0.980 & 263.5\\
20.8 & 255 & 4 & 0.980 & 250.2\\
20.9 & 268 & 7 & 0.980 & 265.5\\
21.0 & 284 & 6 & 0.980 & 281.8\\
21.1 & 287 & 7 & 0.980 & 287.9\\
21.2 & 298 & 8 & 0.980 & 300.1\\
21.3 & 307 & 6 & 0.980 & 309.3\\
21.4 & 322 & 6 & 0.980 & 324.7\\
21.5 & 319 & 5 & 0.979 & 321.7\\
21.6 & 354 & 5 & 0.979 & 356.5\\
21.7 & 408 & 5 & 0.979 & 408.8\\
21.8 & 458 & 8 & 0.979 & 459.0\\
21.9 & 503 & 8 & 0.979 & 505.2\\
22.0 & 494 & 7 & 0.978 & 499.3\\
22.1 & 530 & 4 & 0.977 & 536.4\\
22.2 & 541 & 5 & 0.976 & 550.1\\
22.3 & 538 & 9 & 0.976 & 545.5\\
22.4 & 536 & 9 & 0.974 & 545.0\\
22.5 & 542 & 11 & 0.973 & 553.9\\
22.6 & 532 & 11 & 0.971 & 544.7\\
22.7 & 492 & 12 & 0.969 & 504.7\\
22.8 & 506 & 10 & 0.967 & 517.3\\
22.9 & 487 & 8 & 0.964 & 500.1\\
23.0 & 483 & 6 & 0.961 & 494.8\\
23.1 & 494 & 13 & 0.957 & 510.3\\
23.2 & 465 & 11 & 0.952 & 485.5\\
23.3 & 439 & 8 & 0.946 & 459.9\\
23.4 & 426 & 9 & 0.940 & 444.3\\
23.5 & 385 & 13 & 0.932 & 405.2\\
23.6 & 361 & 9 & 0.923 & 383.9\\
23.7 & 366 & 7 & 0.912 & 388.2\\
23.8 & 350 & 6 & 0.901 & 379.3\\
23.9 & 324 & 5 & 0.889 & 357.5\\
24.0 & 316 & 6 & 0.875 & 354.0\\
24.1 & 310 & 5 & 0.860 & 351.5\\
24.2 & 289 & 9 & 0.841 & 335.5\\
24.3 & 241 & 6 & 0.818 & 287.5\\
24.4 & 239 & 8 & 0.783 & 298.2\\
24.5 & 194 & 5 & 0.724 & 263.9\\
\hline	    
\end{tabular}
\end{table} 
 	    
\section{Comparison with GGC}

\begin{figure}
 \includegraphics[width=8.7cm]{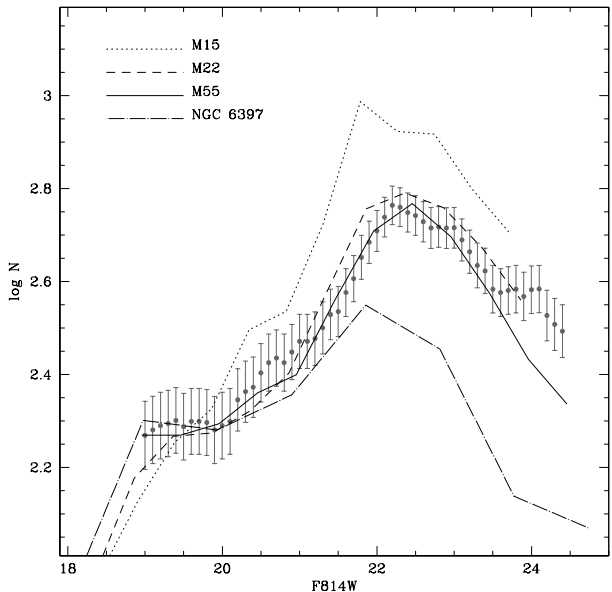}
\caption{Comparison between the F814W LF of $\omega$ Cen (grey points) and the LFs of M15 (dotted
line), NGC 6397 (dot-dashed line), M22 (dashed line) and M55 (solid line).}
\label{gc}
\end{figure}

In Fig. \ref{gc} we compare the deep MS-LF of $\omega$ Cen 
(obtained through the F814W filter) with those calculated by
Piotto \& Zoccali (1999) and Piotto, Cool \& King (1997) for four other Galactic
GCs, namely M15, M22, M55 and NGC 6397. 

To compare the five  LFs we assumed the distance and reddening scale described in
Sollima et al. (2006), the  extinction coefficient $A_{F814W}=1.825~E(B-V)$
and the photometric conversions by Sirianni et al. (2005).  All of the LFs have been
normalized to have the same number of stars in the magnitude range
$19<I_{F814W}<20$ .

The LF of $\omega$ Cen has a slope similar to those of M22 and
M55, being located between the two extreme cases of M15 and NGC 6397. In
particular, M15 shows a significant overabundance of faint stars, incompatible
with the measurement errors. A similar behaviour is observable also in GCs like
M30 and M92 which have a MS-LF  similar to that of M15 (see Piotto \& Zoccali 1999).

Note that Piotto \& Zoccali (1999) measured the LFs close to the cluster
half-mass radius, where mass segregation effects are expected to be less severe.
  
\section{Mass Function}

\begin{figure}
 \includegraphics[width=8.7cm]{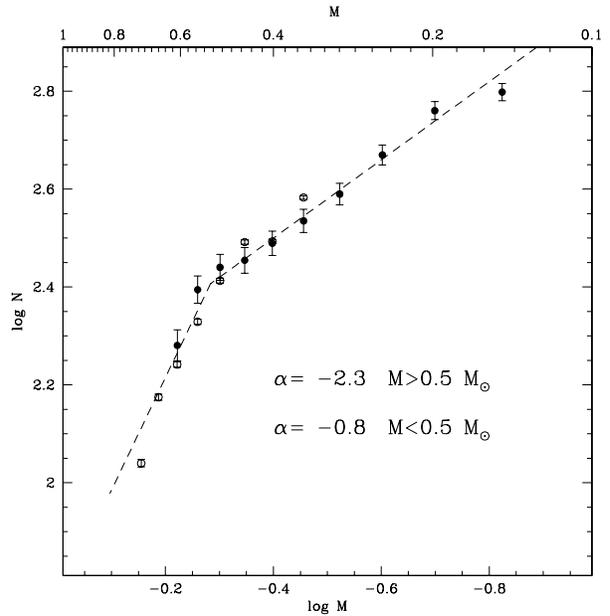}
\caption{Mass Function of $\omega$ Cen calculated from the FORS1 sample at distances $>$12'
from the cluster center (open points) and from the ACS external field (black points). 
A broken power-law with indices $\alpha_{1}=-2.3$ (for $M>0.5~M_{\odot}$) and 
$\alpha_{2}=-0.85$ (for $M<0.5~M_{\odot}$) is overplotted.}
\label{mfshow}
\end{figure}

The LFs shown in Fig. \ref{lfors} and \ref{lfacs} have been converted to MF using the
mass-luminosity relation provided by Baraffe et al. (1997). To convert colors
and magnitudes in the absolute system we adopted the distance modulus $(m-M)_{0}=13.72$ (Sollima
et al. 2006), the reddening $E(B-V)=0.11$ (Lub 2001) and the extinction coefficients $A_{R}=2.35~E(B-V)$ (Savage
\& Mathis 1979) and $A_{F814W}=1.825~E(B-V)$ (Sirianni et al. 2005). 
Considering the negligible effects of dynamical evolution, the averaged LF of the FORS1 sample has been
calculated using all the stars located at distances $r>12'$ from the cluster center. 
In Fig. \ref{mfshow} the calculated MFs are shown. The two MFs have been normalized to have the same number of 
stars in the mass range $0.35 M_{\odot}<M<0.65 M_{\odot}$ . As can be
noted, the obtained MFs agree quite well in the overlap region. 
The MF shown in Fig. \ref{mfshow} presents a well defined broken power-law
shape, with a slope $\alpha \sim-2.3$ for masses $M>0.5M_{\odot}$ and a shallower slope
($\alpha \sim -0.8$) for smaller masses. A similar behaviour of the MF has been found by 
Reid \& Gizis (1997) from the analysis of a sample of Galactic disk stars\footnote{We refer to the power-law 
fit made by these authors in the mass range $0.08 M_{\odot}<M< 0.5 M_{\odot}$ using the 
mass-luminosity relation of Baraffe \& Chabrier (1996) similar to the one adopted in this 
work (see Table 5 in Reid \& Gizis 1997).}.

\begin{figure}
 \includegraphics[width=8.7cm]{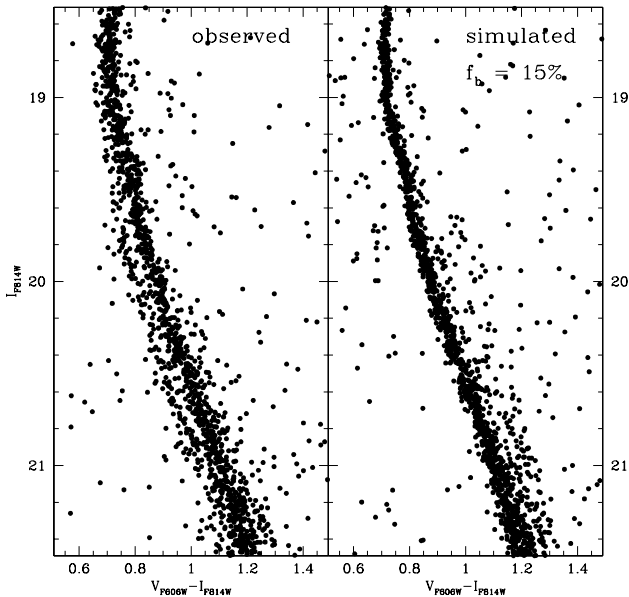}
\caption{Observed ACS CMD (left $panel$) and simulated CMD with a binary fraction $f_{b}=0.15$ 
(right $panel$) in the magnitude range $18.5<I_{F814W}<21.5$.}
\label{binary}
\end{figure}

In principle, two effects can distort the obtained MF:
\begin{itemize}
\item The presence of a significant spread in the metal and
possibly helium content (Norris et al. 1996; Norris 2004) that 
causes significant 
changes in the mass-luminosity relation. 
However, as shown in \S \ref{method}, the shape of the LF is dominated by the metal-poor population of $\omega$ Cen.
For this reason we consider the MF derived here as representative of the dominant cluster population.
\item Unresolved binary systems are shifted in the CMD toward brighter
magnitudes and therefore can distort the derived MF. 
\end{itemize} 
To quantify the impact of a significant binary fraction in the shape of the derived MF, 
we performed a number of CMD simulation with different binary frequencies following 
the prescription of Bellazzini et al. (2002).
The binary population has been simulated by extracting random pairs of 
stars from a broken power-low MF with given indices $\alpha_{1}$ and $\alpha_{2}$. 
The F814W and F606W fluxes of the binary components have been derived using 
the mass-luminosity relation of Baraffe et al. (1997) and
summed in order to obtain the magnitudes of the unresolved binary system.
Field stars were added following the procedure described in \S \ref{field}.    
In Fig. \ref{binary} the observed ACS CMD and the synthetic CMD simulated with a 
fraction of binaries $f_{b}=15\%$ are shown. As expected, a significant number of 
binary systems populate the synthetic CMD in a region located redward with respect to the 
dominant cluster MS ({\it binary region}).   
The comparison of the observed CMD with simulations accounting for a wide range 
of binary fractions indicates that the binary fraction in $\omega$ Cen 
must be smaller than $f_{b} < 15\%$.
Note that at least part of the stars populating the {\it binary region} in the observed 
ACS CMD could be single stars belonging to the metal-rich populations of $\omega$ Cen, 
that are not considered in the simulated CMD. For this reason, the binary fraction estimated 
above represents an upper limit to the true binary fraction in this stellar system.
Then, we derived the LF of the simulated CMD adopting the same procedure described in the 
previous sections in order to quantify the effect of the presence of binary systems.
We found that a broken power-low MF with indices $\alpha_{1}=-2.3$ (for $M>0.5~M_{\odot}$) and 
$\alpha_{2}=-0.8$ (for $M<0.5~M_{\odot}$) reproduces well the observed LF even assuming a 
binary fraction of $f_{b}=15\%$. Therefore, we conclude that binary stars have only 
a negligible effect on the shape of the MF in the considered mass range. 

\begin{figure}
 \includegraphics[width=8.7cm]{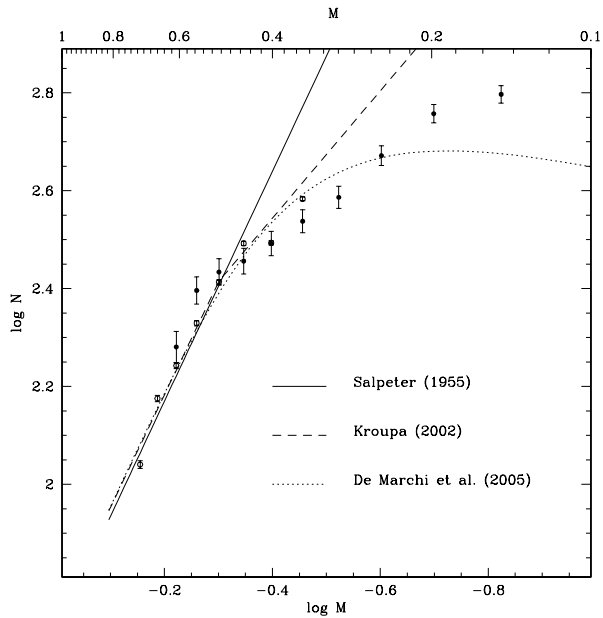}
\caption{Same as Fig. \ref{mfshow}. 
The analytical relation by Salpeter (1955, solid line), De Marchi et al. (2005, dotted line) 
and Kroupa (2002, dashed line) are overplotted.}
\label{mf}
\end{figure}

Some of the analytical IMF listed in Table 1 and the present-day MF derived by De Marchi et al. (2005) 
for a sample of Galactic GCs are overplotted to the MFs obtained in this paper in 
Fig. \ref{mf} . All MFs
have been normalized in the mass range $0.5 M_{\odot}<M<0.8 M_{\odot}$ . 
While for masses $M<0.5 M_{\odot}$ all the analytical IMFs reproduce quite well the observed MF of
$\omega$ Cen, at lower masses the MF of $\omega$ Cen shows a stronger change of slope than those predicted
by the analytical IMFs. In the low-mass range, the MF by De Marchi et al. (2005) predicts a deficiency 
of stars with respect to the MF measured in $\omega$ Cen, as expected in relaxed systems where low-mass stars are lost 
via evaporation and tidal interaction with the Milky Way.  

\section{Discussion}

In Sect. \ref{resforsms} we have shown that the MS-LF measured at different
distances from the center does not show significant modifications, confirming
that $\omega$ Cen is still dynamically young. This result is in agreement
with that found by Anderson (1997) by comparing the observed 
MS-LF of $\omega$ Cen
with the theoretical predictions with and without equipartition and
by Ferraro et al. (2006) on the basis of the comparison between the radial
distribution of the blue straggler stars and {\it normal} less massive
cluster stars. Hence the observed MS-LF (and the derived MF) is expected to be
essentially unaffected by {\it internal} dynamical processes (as e.g. mass 
segregation). Note
that the fact that the MF slope derived for $\omega$ Cen is formally equal
to that measured by Reid \& Gizis (1997) for disk stars in the solar neighborhood 
fully supports this hypothesis. 

A number of works suggests that $\omega$ Cen could be the remnant nucleus 
of a dwarf galaxy which merged in the past with the Milky Way (see Romano et al.
2007 and references therein). In this picture, the cluster experienced strong
tidal losses during its interaction with the Galaxy. 
N-body simulations by Tsuchiya et al. (2004) suggest that the system lost 
$\sim$ 90\% of its initial mass during the first 2 Gyr.
Evidence that seems to
confirm this hypothesis comes from the detection of a  significant stellar overdensity
resembling a pair of tidal tails surrounding $\omega$ Cen (Leon et al. 2000).
However, this result has been questioned by Law et al. (2003) who found that
Leon's et al. tidal tails were strongly correlated with inhomogeneities in the
reddening distribution. 
Note however that the lack of equipartition in
$\omega$ Cen should lead the system to lose stars independently on their masses. 
Therefore, even strong stellar losses should not significantly distort the MS-LF
of the cluster. Hence, the MS-LF shown in Fig. \ref{lfacs} should reflect the 
global primordial luminosity distribution of MS stars in the cluster.
 
If this consideration is true, the 
comparison of the MS-LF shown in Figure 7 casts
some doubts on the "universality" of the IMF. In fact, under the assumption that
all stellar systems formed their stars following a "universal" IMF, we would
expect to observe a general agreement in the MS-LF shape (with respect to that
measured in $\omega$ Cen) in poorly dynamically evolved clusters or a deficiency of 
faint (low-mass) stars in highly evolved clusters where dynamical effects have played 
a significant role.
However in no cases we would expect to see an excess of low-mass stars.  

Indeed, as shown in Figure 7, the MS-LF of $\omega$ Cen seems to share the same
shape as that observed in M22 and M55, but significant differences in the
low-luminosity star content are apparent with respect to M15 and NGC6397. In
particular, while the overabundance of faint stars with respect to NGC6397 can be
interpreted in terms of systematic evaporation of low-mass stars in the highly
evolved cluster NGC6397, the difference with respect to M15 is much more 
puzzling. Indeed,
if the LF in $\omega$ Cen reflects the primordial LF of the cluster, the
difference with respect to M15 could be interpreted only in terms of a "real" 
difference in the IMF. It is worth of noticing that 
other two clusters in the Piotto \& Zoccali (1999) sample (M30 and M92) share the
same behaviour of M15, showing a significant overabundance of faint low-mass
stars with respect to $\omega$ Cen. The slopes of the low-mass end of
the MF derived in these clusters by Piotto \& Zoccali (1999) ($\alpha= -1.1, -1.2~\mbox{and}~
-0.9$ for M15, M30 and M92, respectively) turn out to be smaller
than that measured in this paper for $\omega$ Cen ($\alpha=-0.8$). 
Interestingly enough, all of these clusters have a 
metallicity significantly lower ($[Fe/H]\sim -2$) than the mean 
metallicity of $\omega$ Cen ($[Fe/H]\sim-1.7$, Suntzeff \& Kraft 1996).

This evidence might suggest 
that a "primordial" difference in the IMF of stellar systems as a function of 
the metallicity (i.e. metal poor clusters tend to produce more low-mass stars) 
could exist. The
physical reason for this might be found in the higher efficiency of the
fragmentation process in low-metallicity proto-cluster clouds (see Silk 1977). 
A possible dependence of the slope of the faint end of the clusters MFs on
metallicity was also discussed and not excluded by Piotto \& Zoccali (1999).
Note that a similar result was also presented by McClure et al. (1986) and 
Djorgovski, Piotto \& Capaccioli (1993) from the analysis of the LF of several
Galactic GCs.

Clearly a deeper investigation is required to finally address this important
question. However, the evidence presented here supports the possibility
that the MF measured in $\omega$ Cen could be a reasonable approximation of
the cluster IMF and that a variation of the IMF slope with the cluster
metallicity could exist. 

\section*{acknowledgements}
This research was supported by the Ministero dell'Istruzione, Universit\`a e Ricerca 
and the Agenzia Spaziale Italiana. 
We warmly thank Paolo Montegriffo for assistance during catalogs 
cross-correlation. We also thank Elena Pancino, Cristian Vignali and the anonymous referee for their helpful 
comments and suggestions.

\label{lastpage}

\end{document}